# Limitation of Finite Difference Scheme in Electroconvection with Unipolar Charge Injection: A base-state Analysis

Yifei Guan[1], Junyu Huang[2], and Jian Wu[2] [§]

[1]*Department of Mechanical Engineering, Rice University, Houston, U.S.A. 77005*

[2] *Key Laboratory of Aerospace Thermophysics, School of Energy Science and Engineering, Harbin Institute of Technology, Harbin 150001, PR China*

The 1D hydrostatic base state of electroconvection driven by unipolar charge injection between two parallel electrodes is investigated using a finite difference method. A boundary layer near the anode surface is derived analytically. The computational grid is required to resolve this boundary layer to maintain high order accuracy.

## I. INTRODUCTION

Electroconveciton (EC) phenomenon was first reported by G. I. Taylor in 1966, describing cellular convection in the liquid droplet [1]. Since then, EC has been observed in a number of systems where the interaction of electrostatic force with fluids is present. In nonequilibrium electrohydrodynamic (EHD) systems [1-22], poorly conductive leaky dielectric fluid acquire unipolar charge injection at the surface interface in response to the electric field.

To gain insight into the complexity of the EC flow, the problem can be investigated using numerical simulations. The earlier finite-difference simulations have shown that strong numerical diffusivity may contaminate the model [4]. Other numerical approaches include the particle-in-cell method [23], finite volume method with the flux-corrected transport scheme [24], total variation diminishing scheme [9, 11, 15-17], and the Lattice Boltzmann method [19-22, 25-27]. For any numerical simulations or linear stability analysis [13] of the EC system, a 1D hydrostatic base state (conduction state) has to be obtained a priori before any perturbation is applied. Therefore, the numerical solution of the 1D base state is critical to the performance of more complicated numerical simulations.

In this paper, we performed error analyses of the finite difference method applied to solve the 1D hydrostatic base state of the EC problem. A boundary layer is derived depending on the charge injection level. We found that the boundary layer thickness limits the first grid layer at the anode boundary surface.

## II. GOVERNING EQUATIONS AND DIMENSIONAL ANALYSIS

The governing equations describing the 1D hydrostatic base-state of EC flow include the drift-diffusion equation for charge density and the Poisson equation for electric potential.

---

[§] *jian.wu@hit.edu.cn*

$$\frac{\partial \rho_c^*}{\partial t^*} + \nabla \cdot \left[ \mu_b \rho_c^* E^* - D_c \nabla \rho_c^* \right] = 0, \quad (1)$$

$$\nabla^{*2} \varphi^* = -\frac{\rho_c^*}{\varepsilon}, \quad (2)$$

$$E^* = -\nabla^* \varphi^*, \quad (3)$$

where $\mu_b$ is the ion mobility, $D_c$ is the ion diffusivity, $\rho_c^*$ is the charge density, $\varepsilon$ is the electric permittivity, $\varphi^*$ is the electric potential, $E^*$ is the electric field. In 1D base-state, $E^*$ has the same direction as the drift direction, and thus it also represents the intensity. The variables to be solved are the charge density -- $\rho_c^*$, electric potential -- $\varphi^*$, and electric field intensity – $E^*$. The asterisk * denotes dimensional variables, and the 1D direction is in the x-axis.

In the hydrostatic base-state, the system can be non-dimensionalized with the electric field properties, i.e., $H$ is the distance between the electrodes, $\rho_0$ is the injected charge density at the anode, and $\Delta\varphi_0$ is the voltage difference applied to the electrodes. Respectively, the time $t$ is non-dimensionalized by $H^2/(\mu_b \Delta\varphi_0)$, the charge density $\rho_c^*$ by $\rho_0$, electric potential $\varphi^*$ by $\Delta\varphi_0$, and electric field intensity $E^*$ by $\Delta\varphi_0/H$. Therefore, a non-dimensional form of the governing equations (Eq.(1)-(3)) is:

$$\frac{\partial \rho_c}{\partial t} + \nabla \cdot \left[ \rho_c E - \frac{1}{Fe} \nabla \rho_c \right] = 0, \quad (4)$$

$$\nabla^2 \varphi = -C \rho_c, \quad (5)$$

$$E = -\nabla \varphi, \quad (6)$$

where the $\rho_c$, $\varphi$, and $E$ are the corresponding non-dimensional variables. These non-dimensional governing equations yield two dimensionless parameters describing the system's state [9-22].

$$C = \frac{\rho_0 H^2}{\varepsilon \Delta\varphi_0}, \quad Fe = \frac{\mu_b \Delta\varphi_0}{D_e}, \quad (7)$$

The physical interpretations of these parameters are as follows: $C$ is the charge injection level [13, 18] and $Fe$ is the reciprocal of the charge diffusivity coefficient [13, 18].

In general, Eq. (4)-(6) cannot be solved analytically. In practice, however, the charge diffusion term is much smaller than the drift term. Therefore, to perform accuracy analysis, we assume $Fe \to \infty$ and Eq. (4)-(6) are further reduced to:

$$\frac{\partial \rho_c}{\partial t} + \nabla \cdot (\rho_c E) = 0, \quad (8)$$

$$\nabla^2 \varphi = -C \rho_c, \quad (9)$$

$$E = -\nabla \varphi, \quad (10)$$

In steady-state ($\frac{\partial \rho_c}{\partial t} = 0$), the analytical solutions of charge density $\rho_c$, electric potential $\varphi$, and electric field intensity $E$ can be obtained from Eq. (8)-(10).

$$\rho_c^{analytical} = \frac{a}{C}(x+b)^{-1/2}, \tag{11}$$

$$\varphi^{analytical} = -\frac{4}{3}a(x+b)^{3/2} + c, \tag{12}$$

$$E^{analytical} = 2a(x+b)^{1/2}, \tag{13}$$

where $a$, $b$, and $c$ are constants depending on the boundary conditions and the charge injection level $C$. To solve for these three unknown constants, we need three equations from the boundary conditions. The boundary conditions used in EHD modelings are Dirichlet boundary conditions for $\varphi$ on both electrodes and Dirichlet boundary conditions for $\rho_c$ on the anode. $E$ can be solved from the first derivative of $\varphi$. **Table I** summarizes the boundary conditions for the EHD system.

**Table I.** Boundary conditions for the EHD system described by Eq. (11)-(13)

| Boundary | Charge density | Electric potential |
|---|---|---|
| Anode (x=0) | $\rho_c(x=0) = 1$ | $\varphi(x=0) = 1$ |
| Cathode (x=1) | Determined by Eq. (8) | $\varphi(x=1) = 0$ |

where $x = x^*/H$ is the non-dimensional spatial coordinate. Therefore, the constants $a$, $b$, and $c$ can be determined by:

$$\frac{4}{3}a\left[(1+b)^{3/2} - b^{3/2}\right] = 1, \tag{14}$$

$$b = \left(\frac{a}{C}\right)^2, \tag{15}$$

$$c = 1 + \frac{4}{3}ab^{3/2}, \tag{16}$$

We discuss three scenarios when C=1, 10, and 100, corresponding to weak, medium, and strong charge injection respectively.

**Table II.** Constants $a$, $b$, and $c$ of different injection levels $C$

| Constants | C=1 | C=10 | C=100 |
|---|---|---|---|
| a | 0.56278 | 0.74412 | 0.74994 |
| b | 0.31672 | 0.0055371 | 0.000056241 |
| c | 1.13375 | 1.0004 | 1.0000 |

**FIG. 1** shows the analytical solution of charge density $\rho_c$ and electric field E for three charge injection levels. The initial decay of the $\rho_c$ profile is highly dependent on the length scale b, which can be regarded as a boundary layer of the reduced system Eq. (8)-(10).

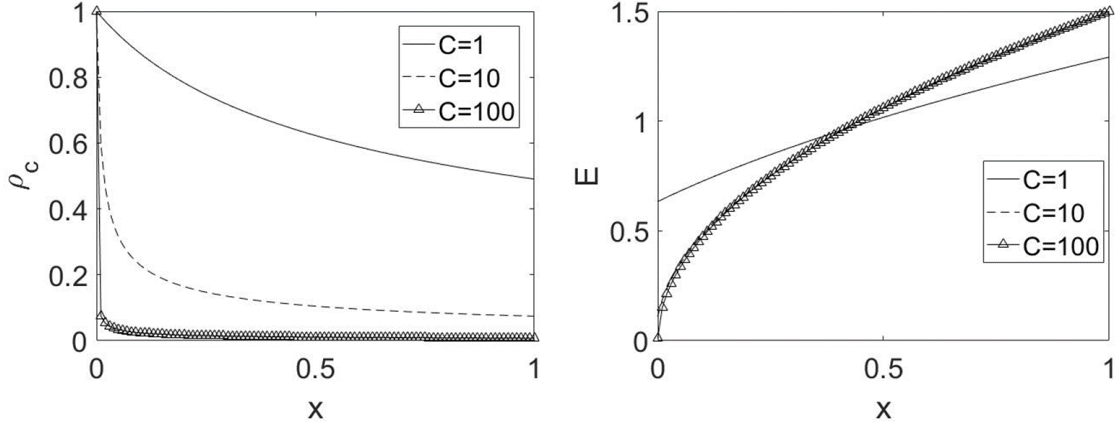

**FIG. 1.** Analytical solution of $\rho_c$ (left) and $E$ (right) for weak, medium, and strong charge injection level.

### III. ANALYTICAL ANALYSIS

To study the accuracy of the finite difference scheme applied to the EHD problem, we perform Taylor expansion on Eq. (11)-(13) at $x_j = j\Delta x$, where $j = 0, 1, ..., 1/\Delta x$:

$$\rho_c^{analytical}\left(x_j + \Delta x\right) = \frac{a}{C}\left(\Delta x + x_j + b\right)^{-1/2}$$
$$= \frac{a}{C(x_j+b)^{1/2}} - \frac{a\Delta x}{2C(x_j+b)^{3/2}} + \frac{3a\Delta x^2}{8C(x_j+b)^{5/2}} + O(\Delta x^3), \quad (17)$$

$$\varphi^{analytical}\left(x_j + \Delta x\right) = -\frac{4}{3}a\left(\Delta x + x_j + b\right)^{3/2} + c$$
$$= c - \frac{4}{3}a(x_j+b)^{3/2} - 2\Delta x a(x_j+b)^{1/2} - \frac{a\Delta x^2}{2(x_j+b)^{1/2}} + O(\Delta x^3), \quad (18)$$

$$E^{analytical}\left(x_j + \Delta x\right) = 2a\left(\Delta x + x_j + b\right)^{1/2}$$
$$= 2a(x_j+b)^{1/2} + \frac{a\Delta x}{(x_j+b)^{1/2}} - \frac{a\Delta x^2}{4(x_j+b)^{3/2}} + O(\Delta x^3), \quad (19)$$

The one-sided second order accurate finite difference scheme at the anode boundary $x_j = 0$ derived from Eq. (17)-(19) yields:

$$\left.\frac{d\rho_c}{dx}\right|_{x=0}^{numerical} \approx \frac{-3\rho_c(0)+4\rho_c(\Delta x)-\rho_c(2\Delta x)}{2\Delta x} \tag{20}$$

$$= -\frac{a}{2C(b)^{3/2}} + O(\Delta x^2) = \left.\frac{d\rho_c}{dx}\right|_{x=0}^{analytical} + O(\Delta x^2)$$

$$\left.\frac{d\varphi}{dx}\right|_{x=0}^{numerical} \approx \frac{-3\varphi(0)+4\varphi(\Delta x)-\varphi(2\Delta x)}{2\Delta x} \tag{21}$$

$$= -2ab^{1/2} + O(\Delta x^2) = \left.\frac{d\varphi}{dx}\right|_{x=0}^{analytical} + O(\Delta x^2)$$

$$\left.\frac{dE}{dx}\right|_{x=0}^{numerical} \approx \frac{-3E(0)+4E(\Delta x)-E(2\Delta x)}{2\Delta x} \tag{22}$$

$$= ab^{-1/2} + O(\Delta x^2) = \left.\frac{dE}{dx}\right|_{x=0}^{analytical} + O(\Delta x^2)$$

The second derivative can be derived in a similar manner. The higher-order terms $O(\Delta x^n) = O(\Delta x^n b^{-m})$, for n>0 and m>0. Since $|n-m| < \infty$, $\lim_{n,m\to\infty} \Delta x^n b^{-m} \to 0$ or $O(\Delta x^n b^{-m})$ converges for $\Delta x < b$. Therefore, it requires $\Delta x < b$ at the anode boundary ($j=0$) for the finite difference scheme to converge. Otherwise, the numerical error generated at this boundary will be unexpected and may contaminate the whole system. The length scale b can be regarded as a boundary layer of EC at the anode surface, and it needs to be resolved.

## IV. NUMERICAL RESULTS

The numerical scheme contains center difference for the interior grid points and one-sided forward or backward difference for the boundary grid points. The first and second derivative matrices are given as:

$$D_1 = \frac{1}{2\Delta x}\begin{bmatrix} -3 & 4 & -1 & & & \\ -1 & & 1 & & & \\ & -1 & \dots & 1 & & \\ & & -1 & & 1 & \\ & & & 1 & -4 & 3 \end{bmatrix}, \quad D_2 = \frac{1}{\Delta x^2}\begin{bmatrix} 2 & -5 & 4 & -1 & & \\ 1 & -2 & 1 & & & \\ & 1 & \dots & 1 & & \\ & & 1 & -2 & 1 & \\ & & -1 & 4 & -5 & 2 \end{bmatrix}, \tag{23}$$

where $D_1$ and $D_2$ are first and second-order finite difference matrices and $\Delta x$ is the grid spacing.

To perform error analysis, we use the analytical solution of either $\rho_c$ or $\varphi$ to compute the numerical solutions of the other two variables with the finite difference scheme. When $\rho_c = \rho_c^{analytical}$, the numerical solutions of $\varphi$ and $E$ can be obtained by solving the following linear equations:

$$D_2 \varphi^{\rho_c} = -C\rho_c^{analytical}, \tag{24}$$

$$E^{\rho_c} = -D_1 \varphi^{\rho_c}, \tag{25}$$

When $\varphi = \varphi^{analytical}$, the numerical solutions of $\rho_c$ and $E$ can be obtained by solving the following linear equations:

$$\rho_c^\varphi = -\frac{1}{C} D_2 \varphi^{analytical}, \tag{26}$$

$$E^\varphi = -D_1 \varphi^{analytical}, \tag{27}$$

The superscript denotes the variable of its analytical value. The error is defined by the $L_2$ norm for resolved variable $\gamma$ [28]:

$$Er = \sqrt{\Delta x \sum \left(\gamma^{analytical} - \gamma^{numerical}\right)^2}, \tag{28}$$

A numerical error analysis is shown in **FIG. 2** and **FIG. 3**. For C=1 (b=0.32), the finite difference scheme exhibits second-order accuracy as expected from construction. For larger C (smaller b), however, the accuracy degrades for $\Delta x > b$, which is consistent with the analysis based on Taylor expansion. Among the resolved variables, b has the most significant effect on the error of $\rho_c$ because b has the greatest power at the denominator of the equations of $\rho_c$ (Eq. (17)).

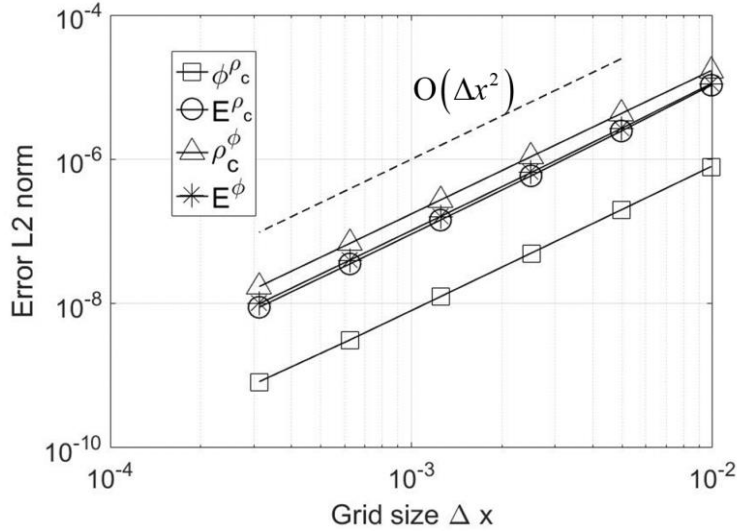

**FIG. 2.** Error with respect to grid size. For C=1, the finite difference exhibits second-order accuracy for $\Delta x < 10^{-2}$.

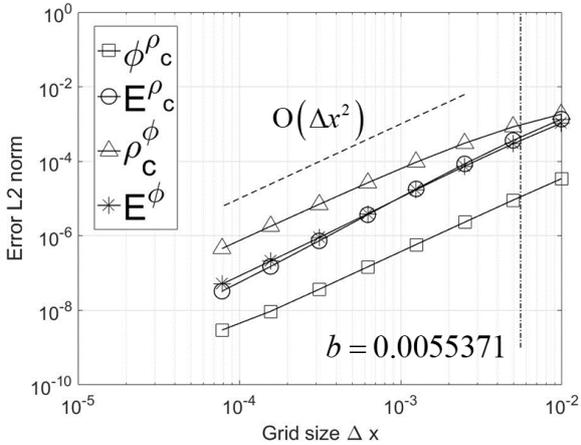
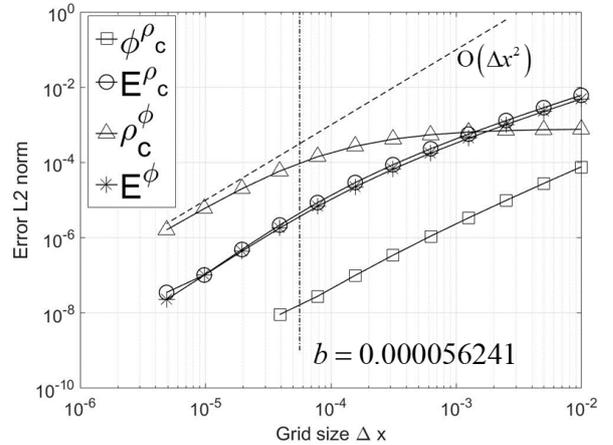

**FIG. 3.** For C=10 (left) and C=100 (right), the numerical scheme is inaccurate for large $\Delta x$. Only for $\Delta x < b$, the finite difference method approaches second-order accuracy.

To validate the error analysis in the coupled 1D system, the Eq. (8)-(10) can be solved numerically. The spatial derivative is approximated by the finite difference scheme using the derivative matrices in Eq. (28) and the temporal stepping scheme is the two-step Adams Bashforth scheme. Steady-state is obtained when the maximum difference of $\rho_c$ at sequential time step is less than $10^{-14}$. **FIG. 4** shows the error analysis for the coupled 1D system at C=100 (strong charge injection). The error is significant for large grid sizes $\Delta x$, and the numerical scheme approaches second-order accuracy when $\Delta x < b$. The numerical result is consistent with the analysis based on Taylor expansion.

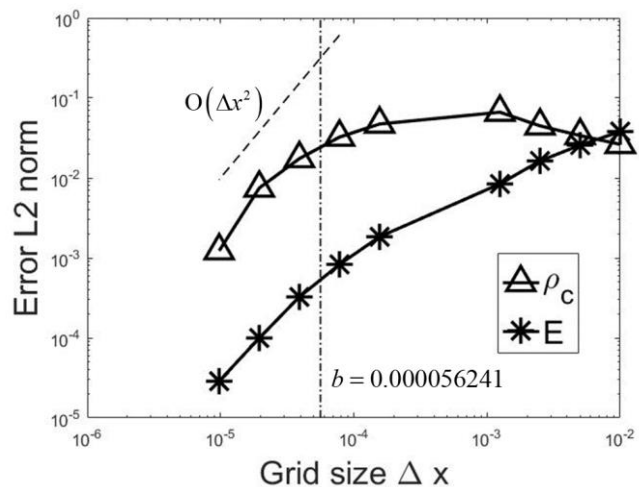

**FIG. 4.** The numerical simulation is performed on the coupled 1D system at C=100. The numerical scheme is inaccurate for large $\Delta x$. Only for $\Delta x < b$, the finite difference method approaches second-order accuracy.

## V. CONCLUSION

The 1D hydrostatic base state of electroconvection (EC) is solved using a second-order finite difference method. Error analysis is performed using a Taylor series expansion and numerical simulation. To maintain second order accuracy, the grid size at the anode boundary is constrained depending on the charge injection level, i.e., $\Delta x < b$, where b can be regarded as an EC boundary layer. Although a unipolar discharge is assumed in EC simulation and no physical layer (for example, electric double layer) is considered, the EC boundary layer b has to be resolved in order to maintain high accuracy. The proposed condition can easily be satisfied by using a stretched grid, and it can be applied to all Taylor expansion based numerical schemes for EC simulation.

## VI. REFERENCES

[1] G. I. Taylor, Studies in electrohydrodynamics. I. The circulation produced in a drop by an electric field, Proceedings of the Royal Society of London. Series A. Mathematical and Physical Sciences **291**, 159 (1966).

[2] D. C. Jolly and J. R. Melcher, Electroconvective instability in a fluid layer, Proceedings of the Royal Society of London. A. Mathematical and Physical Sciences **314**, 269 (1970).
[3] B. Malraison and P. Atten, Chaotic behavior of instability due to unipolar ion injection in a dielectric liquid, Physical Review Letters **49**, 723 (1982).
[4] A. Castellanos and P. Atten, Numerical modeling of finite amplitude convection of liquids subjected to unipolar injection, IEEE transactions on industry applications, 825 (1987).
[5] Z. A. Daya, V. B. Deyirmenjian, and S. W. Morris, Bifurcations in annular electroconvection with an imposed shear, Physical Review E **64**, 036212 (2001).
[6] K. Adamiak and P. Atten, Simulation of corona discharge in point–plane configuration, Journal of electrostatics **61**, 85 (2004).
[7] P. Tsai, Z. A. Daya, and S. W. Morris, Aspect-ratio dependence of charge transport in turbulent electroconvection, Physical review letters **92**, 084503 (2004).
[8] P. Tsai, Z. A. Daya, and S. W. Morris, Charge transport scaling in turbulent electroconvection, Physical Review E **72**, 046311 (2005).
[9] P. Traoré and A. Pérez, Two-dimensional numerical analysis of electroconvection in a dielectric liquid subjected to strong unipolar injection, Physics of Fluids **24**, 037102 (2012).
[10] P. Traoré and J. Wu, On the limitation of imposed velocity field strategy for Coulomb-driven electroconvection flow simulations, Journal of Fluid Mechanics **727** (2013).
[11] J. Wu, P. Traoré, P. A. Vázquez, and A. T. Pérez, Onset of convection in a finite two-dimensional container due to unipolar injection of ions, Physical Review E **88**, 053018 (2013).
[12] A. Pérez, P. Vázquez, J. Wu, and P. Traoré, Electrohydrodynamic linear stability analysis of dielectric liquids subjected to unipolar injection in a rectangular enclosure with rigid sidewalls, Journal of Fluid Mechanics **758**, 586 (2014).
[13] M. Zhang, F. Martinelli, J. Wu, P. J. Schmid, and M. Quadrio, Modal and non-modal stability analysis of electrohydrodynamic flow with and without cross-flow, Journal of Fluid Mechanics **770**, 319 (2015).
[14] P. Traore, J. Wu, C. Louste, P. A. Vazquez, and A. T. Perez, Numerical study of a plane poiseuille channel flow of a dielectric liquid subjected to unipolar injection, IEEE Transactions on Dielectrics and Electrical Insulation **22**, 2779 (2015).
[15] J. Wu and P. Traoré, A finite-volume method for electro-thermoconvective phenomena in a plane layer of dielectric liquid, Numerical Heat Transfer, Part A: Applications **68**, 471 (2015).
[16] J. Wu, A. T. Perez, P. Traore, and P. A. Vazquez, Complex flow patterns at the onset of annular electroconvection in a dielectric liquid subjected to an arbitrary unipolar injection, IEEE Transactions on Dielectrics and Electrical Insulation **22**, 2637 (2015).
[17] J. Wu, P. Traoré, A. T. Pérez, and P. A. Vázquez, On two-dimensional finite amplitude electroconvection in a dielectric liquid induced by a strong unipolar injection, Journal of Electrostatics **74**, 85 (2015).
[18] M. Zhang, Weakly nonlinear stability analysis of subcritical electrohydrodynamic flow subject to strong unipolar injection, Journal of Fluid Mechanics **792**, 328 (2016).
[19] K. Luo, J. Wu, H.-L. Yi, and H.-P. Tan, Lattice Boltzmann model for Coulomb-driven flows in dielectric liquids, Physical Review E **93**, 023309 (2016).
[20] K. Luo, J. Wu, H.-L. Yi, and H.-P. Tan, Three-dimensional finite amplitude electroconvection in dielectric liquids, Physics of Fluids **30**, 023602 (2018).
[21] K. Luo, J. Wu, H.-L. Yi, L.-H. Liu, and H.-P. Tan, Hexagonal convection patterns and their evolutionary scenarios in electroconvection induced by a strong unipolar injection, Physical Review Fluids **3**, 053702 (2018).

[22] K. Luo, T.-F. Li, J. Wu, H.-L. Yi, and H.-P. Tan, Mesoscopic simulation of electrohydrodynamic effects on laminar natural convection of a dielectric liquid in a cubic cavity, Physics of Fluids **30**, 103601 (2018).

[23] R. Chicón, A. Castellanos, and E. Martin, Numerical modelling of Coulomb-driven convection in insulating liquids, Journal of Fluid Mechanics **344**, 43 (1997).

[24] P. Vazquez, G. Georghiou, and A. Castellanos, Characterization of injection instabilities in electrohydrodynamics by numerical modelling: comparison of particle in cell and flux corrected transport methods for electroconvection between two plates, Journal of Physics D: Applied Physics **39**, 2754 (2006).

[25] Y. Guan and I. Novosselov, Two Relaxation Time Lattice Boltzmann Method Coupled to Fast Fourier Transform Poisson Solver: Application to Electroconvective Flow, Journal of Computational Physics **397**, 108830 (2019).

[26] Y. Guan and I. Novosselov, Numerical analysis of electroconvection in cross-flow with unipolar charge injection, Physical Review Fluids **4**, 103701 (2019).

[27] Y. Guan, J. Riley, and I. Novosselov, Three-dimensional electroconvective vortices in cross flow, Physical Review E **101**, 033103 (2020).

[28] R. J. LeVeque, *Finite difference methods for ordinary and partial differential equations: steady-state and time-dependent problems* (Siam, 2007), Vol. 98.